\documentclass[aps,prl,reprint,twocolumn,superscriptaddress]{revtex4-2} 
\usepackage{amsfonts}
\usepackage{amssymb}
\usepackage{amsthm,amsmath}
\usepackage{mathrsfs}
\usepackage{indentfirst}
\usepackage{multirow}
\usepackage{graphicx}
\usepackage{amsmath}
\usepackage[amssymb]{SIunits}
\usepackage{amssymb}
\usepackage{color}
\usepackage[usenames,dvipsnames]{xcolor}
\usepackage{natbib}
\usepackage[T1]{fontenc}
\usepackage[hidelinks, bookmarks=true]{hyperref}
\usepackage{bm}
\usepackage{booktabs}
\usepackage{multirow}
\usepackage{setspace}

\begin{document}

\title{Physical Mechanism Behind the Early Onset of the Ultimate State \\in Supergravitational Centrifugal Thermal Convection}

\author{Lei Ren}
\affiliation{New Cornerstone Science Laboratory, Center for Combustion Energy, Key Laboratory for Thermal Science and Power Engineering of MoE, Department of Energy and Power Engineering, Tsinghua University, Beijing 100084, China}%

\author{Jun Zhong}
\affiliation{New Cornerstone Science Laboratory, Center for Combustion Energy, Key Laboratory for Thermal Science and Power Engineering of MoE, Department of Energy and Power Engineering, Tsinghua University, Beijing 100084, China}%

\author{Rushi Lai}
\affiliation{Department of Engineering Mechanics, School of Aerospace Engineering, Tsinghua University, Beijing 100084, China}%

\author{Chao Sun}%
\thanks{chaosun@tsinghua.edu.cn}
\affiliation{New Cornerstone Science Laboratory, Center for Combustion Energy, Key Laboratory for Thermal Science and Power Engineering of MoE, Department of Energy and Power Engineering, Tsinghua University, Beijing 100084, China}%
\affiliation{Department of Engineering Mechanics, School of Aerospace Engineering, Tsinghua University, Beijing 100084, China}%

\date{\today}

\begin{abstract}
We present a combined experimental and numerical investigation of the transition from the classical to the ultimate regime of thermal turbulence in a supergravitational centrifugal convection system. The transition is found to be robust, with the critical Rayleigh number decreasing systematically as the Froude number, defined as the ratio of centrifugal to Earth’s gravity, decreases, highlighting the effect of residual gravity. Once the Rayleigh number reaches the transition threshold, the Stewartson layer induced by residual Earth gravity becomes comparable in thickness to the viscous boundary layer, and their interaction results in a coupled flow that distorts the viscous boundary layer, triggering its transition from laminar to turbulent flow and leading to a sharp increase in heat transport. These findings demonstrate the key role of the Stewartson layer induced by residual gravity in facilitating the transition to the ultimate regime in supergravitational centrifugal thermal convection.
\end{abstract}

\maketitle

Turbulent flows driven by thermal forcing are ubiquitous in nature \cite{Mckenzie1974, Wyngaard1992} and in industrial processes \cite{Owen2015}. A widely used model for investigating such flows is Rayleigh--B\'enard convection (RBC), which consists of a fluid layer confined between two horizontal plates, heated from below and cooled from above \cite{Ahlers2009, Lohse2010, Chilla2012, Lohse2024RMP}. In geophysical and astrophysical systems, the driving strength of these flows reaches extremely high values that cannot be reproduced in laboratory experiments. To bridge this gap, asymptotic scaling laws are often employed to extrapolate laboratory findings to such extreme regimes.
In particular, the relation between the Nusselt number $Nu$ (a dimensionless measure of convective heat transport) and the Rayleigh number $Ra$ (a dimensionless measure of buoyancy driving) provides a powerful framework for exploring these flows. The seminal work of \citet{Kraichnan1962} first predicted that, at sufficiently large $Ra$, turbulent flow enters the so-called ultimate regime, in which the Nusselt number scales as $Nu \sim Ra^{1/2}(\ln Ra)^{-3/2}$. A recent model by \citet{Shishkina2024} for heat transfer in this regime incorporates the dependence on the Prandtl number ($Pr$) and satisfies the mathematically strict upper-bound condition, yielding the scaling relation $Nu \sim Pr^{\pm1/2}Ra^{1/2}(\log Ra)^{-2}$, with a negative exponent $(-1/2)$ for $Pr \ge 1$ and a positive exponent $(+1/2)$ for $Pr \le 1$. Both models predict that the effective scaling exponent $\gamma$ in the relation $Nu \sim Ra^{\gamma}$ exceeds $1/3$ in the ultimate regime—steeper than in the classical regime, where $\gamma < 1/3$—and asymptotically approaches $1/2$ at very large $Ra$.

Over the past 60 years, the existence of the ultimate regime has been the subject of intensive debate and extensive investigation in theory \cite{Malkus1954b, Kraichnan1962, Castaing1989, Shraiman1990, Chavanne1997, Grossmann2011, Grossmann2012, Shishkina2024} and in laboratory experiments \cite{Heslot1987, Castaing1989, Chavanne1997, Glazier1999, Niemela2000, Ahlers2009NJP, Funfschilling2009, He2012, Jiang2020, Jiang2022} (see Ref.~\cite{Lohse2024RMP} for a review). However, in classical RBC systems, where Earth’s gravity is aligned with the temperature gradient, it remains challenging to reach extremely high $Ra$ and to obtain convincing evidence across a wide $Ra$ range \cite{He2012}. To address this limitation, \citet{Jiang2020} introduced a supergravitational centrifugal convection (CC) system, in which centrifugal acceleration acts as an effective gravity parallel to the temperature gradient, while Earth’s gravity is perpendicular to it. The CC system shares lots of similarities with the rotating Rayleigh--B\'enard convection (RRBC) system, which has been widely studied over the past years \cite{Liao2006, King2009, Zhong2009, Kunnen2011, Kunnen2013, King2013, Cheng2018, Zhang2020, Wedi2021, Ecke2022} (see Ref.~\cite{Ecke2023} for a review). However, the key difference is the direction of the rotation axis and the temperature gradient. In RRBC, the rotation axis is aligned with the temperature gradient as well as the buoyancy, in which a strong rotation has a great depression on convection \cite{Ecke2023}, but has little effect on the heat transport in CC \cite{Jiang2020}. Hence, the flow structures, i.e., the boundary zonal flow and the Taylor columns in RRBC, are different from those in CC.

The configuration of CC enables access to much larger $Ra$ and has provided direct evidence for the existence of the ultimate regime over more than a decade in $Ra$ \cite{Jiang2022}. However, a key unresolved question concerns the nature of the transition from the classical to the ultimate regime in the CC system. Specifically, what physical mechanisms govern this transition, and why does it occur at a much lower threshold in the CC system [$Ra^* = \mathcal{O}(10^{10})$] \cite{Jiang2020, Jiang2022} than in the classical RBC system [$Ra^* = \mathcal{O}(10^{12-14})$] \cite{Chavanne1997, He2012, Lohse2024RMP}?

\begin{figure*}[t]
\includegraphics[width=1.0\linewidth]{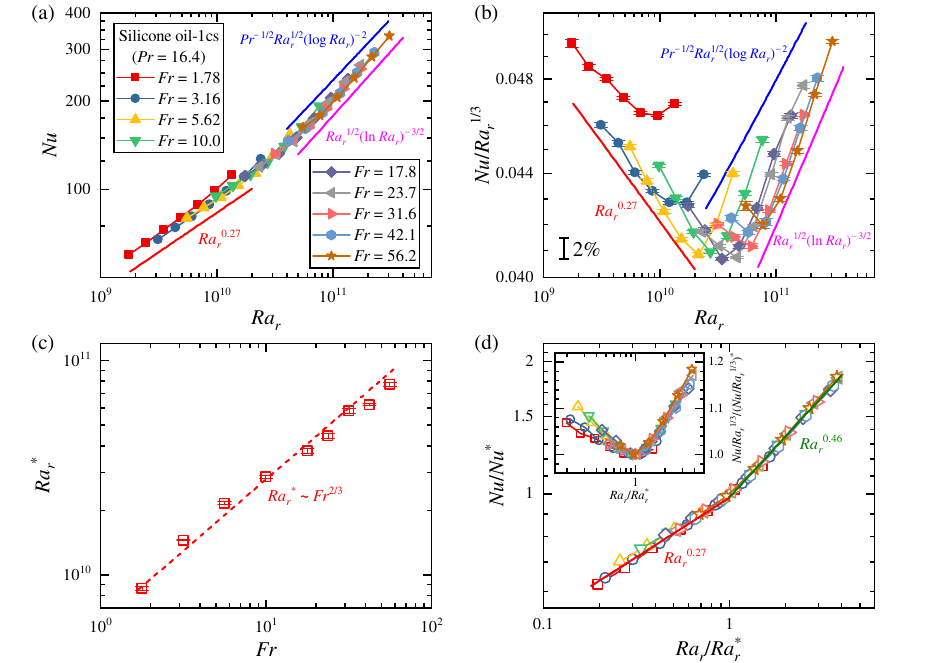}
\caption{\label{Fig1}(a) Measured Nusselt number $Nu$ as a function of Rayleigh number $Ra_r$ for different Froude numbers $Fr$. (b) The compensated plots of $Nu/Ra_r^{1/3}$ versus $Ra_r$. (c) The critical Rayleigh number $Ra_r^*$ for the transition from the classical regime to the ultimate regime as a function of Froude number $Fr$. (d) The normalized $Nu/Nu^*$ versus normalized $Ra_r/Ra_r^*$. Inset: the inset shows the normalized compensated plots. The solid red line in (a), (b), and (d) represents $Nu \sim Ra_r^{0.27}$ based on fitting; the solid blue line in (a) and (b) corresponds to $Nu \sim Pr^{-1/2}Ra_r^{1/2}(\log Ra_r)^{-2}$\cite{Shishkina2024}; the solid pink line in (a) and (b) denotes $Nu \sim Ra_r^{1/2}(\ln Ra_r)^{-3/2}$\cite{Kraichnan1962}; and the solid dark-green line in (d) represents $Nu \sim Ra_r^{0.46}$ based on fitting. The dashed red line in (c) corresponds to $Ra_r^* \sim Fr^{2/3}$.}
\end{figure*}

Recently, \citet{Yao2025} observed an evolution of the flow structure from gravity-dominated vertical convection to centrifugal-dominated RBC as centrifugal buoyancy increased. \citet{Lai2025} investigated the influence of Earth’s gravity on turbulent flow in the CC system and demonstrated that the Stewartson layers \cite{Stewartson1957, Stewartson1966} forming near the inner (cooled) and outer (heated) cylinders are induced by Earth’s gravity. During the transition from the classical to the ultimate regimes, the laminar-type Prandtl–Blasius boundary layer is expected to evolve into the turbulent-type Prandtl–von Kármán boundary layer through the non-normal-nonlinear mechanism \cite{Roche2020, Lohse2023, Lohse2024RMP}. Therefore, in the CC system, the interaction between the Stewartson layers and the viscous boundary layers near the inner (cooled) and outer (heated) cylinders may play a critical role in this transition.

In this work, we experimentally observe that the transition from the classical regime, characterized by an effective scaling of $Nu \sim Ra^{0.27}$, to the ultimate regime, with a much steeper effective exponent $\gamma = 0.46$, is robust in the CC system across a wide range of Froude numbers, which quantify the ratio of centrifugal to Earth’s gravity. Our results further show that the critical Rayleigh number for the onset of the ultimate regime decreases as the Froude number decreases, revealing the influence of residual gravity. By combining experiments and direct numerical simulations with theoretical models, we demonstrate the critical role of Stewartson layers induced by residual gravity in triggering the early onset of the ultimate regime in centrifugal convection, in contrast to the classical RBC system.

\textit{Experimental setup$-$}We adopt the same experimental setup used in previous studies \cite{Jiang2020, Jiang2022}. Since this experimental setup has already been described in detail in \cite{Jiang2020, Jiang2022}, here we only provide some main features. This experimental setup is a cylindrical annulus with a cooled inner wall and a heated outer wall under a solid-body rotation around its vertical axis. The outer (inner) radius of the cooled inner (heated outer) wall is $R_i=120$ mm ($R_o=240$ mm) and the height of the cylindrical annulus is $H=120$ mm, resulting in a gap of $L=R_o-R_i=120$ mm, a radius ratio of $\eta=R_i/R_o=0.5$, and an aspect ratio of $\Gamma=H/L=1$.

The dimensionless governing equations, parameters of the CC system and the numerical methods are presented in the End Matter. To explore the effect of gravity, one can take the Froude number $Fr = \Omega^2 R_m/g$ as another dimensionless parameter that is a measure of the centrifugal acceleration compared to the gravitational acceleration. We use Silicone oil-1cs (Shin-Etsu Chemical Co., Ltd.) as the working fluid to investigate the critical Rayleigh number $Ra_r^*$ for the transition from the classical to the ultimate regime. In the End Matter, the physical properties of Silicone oil-1cs at 25 $^\circ$C are listed in Table~\ref{tab1}, and the parameter space explored in the $Fr$–$Ra_r$ and $Pr$–$Ra_r$ planes is shown in Figs.~\ref{Fig4}(a) and \ref{Fig4}(b), respectively.

\textit{Measurement results$-$}Figure~\ref{Fig1}(a) shows the measured $Nu$ as a function of $Ra_r$ for a wide range of $Fr$, from 1.78 to 56.2. In the low-$Ra_r$ regime, the data follow an effective scaling of $Nu \sim Ra_r^{0.27}$ (solid red line), consistent with the previous study \cite{Jiang2022}, indicating that the system is in the classical regime. In the high-$Ra_r$ regime, the data agree with the theoretical predictions for the ultimate regime, $Nu \sim Pr^{-1/2}Ra_r^{1/2}(\log Ra_r)^{-2}$ \cite{Shishkina2024} (solid blue line) and $Nu \sim Ra_r^{1/2}(\ln Ra_r)^{-3/2}$ \cite{Kraichnan1962} (solid pink line), exhibiting a steep effective scaling exponent within the current parameter range.

For better visibility of the onset, Fig.~\ref{Fig1}(b) shows the plots compensated by $Ra_r^{1/3}$, which correspond to the data in Fig.~\ref{Fig1}(a). The transition from the classical to the ultimate regime is robust across all $Fr$, as evidenced by the change in the effective scaling exponent from $\gamma < 1/3$ to $\gamma > 1/3$. Interestingly, the critical Rayleigh number for this transition, $Ra_r^*$, varies with $Fr$. For example, $Ra_r^*$ is approximately $9\times10^{9}$ for $Fr = 1.78$, while for $Fr = 56.2$, it is close to $8\times10^{10}$. The exact onset of the steeper scaling, corresponding to the ultimate turbulence regime, depends on $Fr$, consistent with the view that the transition is of a non-normal-nonlinear nature \cite{Roche2020, Lohse2023, Lohse2024RMP} and is sensitive to perturbations induced by Earth-gravity-driven flow in the current case.

To determine the critical $Ra_r^*$ for the transition from the classical regime to the ultimate regime, we adopt spline interpolation with $n \ge 3$ data points near the transition on the $Nu/Ra_r^{1/3} \sim Ra_r$ curve in Fig.~\ref{Fig1}(b) and identify the minimum value of each interpolation curve as the $Ra_r^*(n)$ for the corresponding $n$ data points. The critical $Ra_r^*$ is obtained by the mean value of all $Ra_r^*(n)$, and the standard deviation is used as its associated uncertainty. Figure~\ref{Fig1}(c) shows the critical $Ra_r^*$ as a function of $Fr$. Interestingly, the critical $Ra_r^*$ increases with increasing $Fr$, following a power law $Ra_r^* \sim Fr^{2/3}$. Figure~\ref{Fig1}(d) shows the results of normalizing the datasets in Fig.~\ref{Fig1}(a) by the critical $Ra_r^*$  obtained from experiment and the corresponding $Nu^*$. Remarkably, the data for different $Fr$ now collapse onto two master curves: a scaling exponent of 0.27 for $Ra_r/Ra_r^* < 1$ and $Nu \sim Ra_r^{0.46}$ for $Ra_r/Ra_r^* > 1$, clearly demonstrating the universal nature of the scaling relation. 

\begin{figure}[t]
\includegraphics[width=1.0\linewidth]{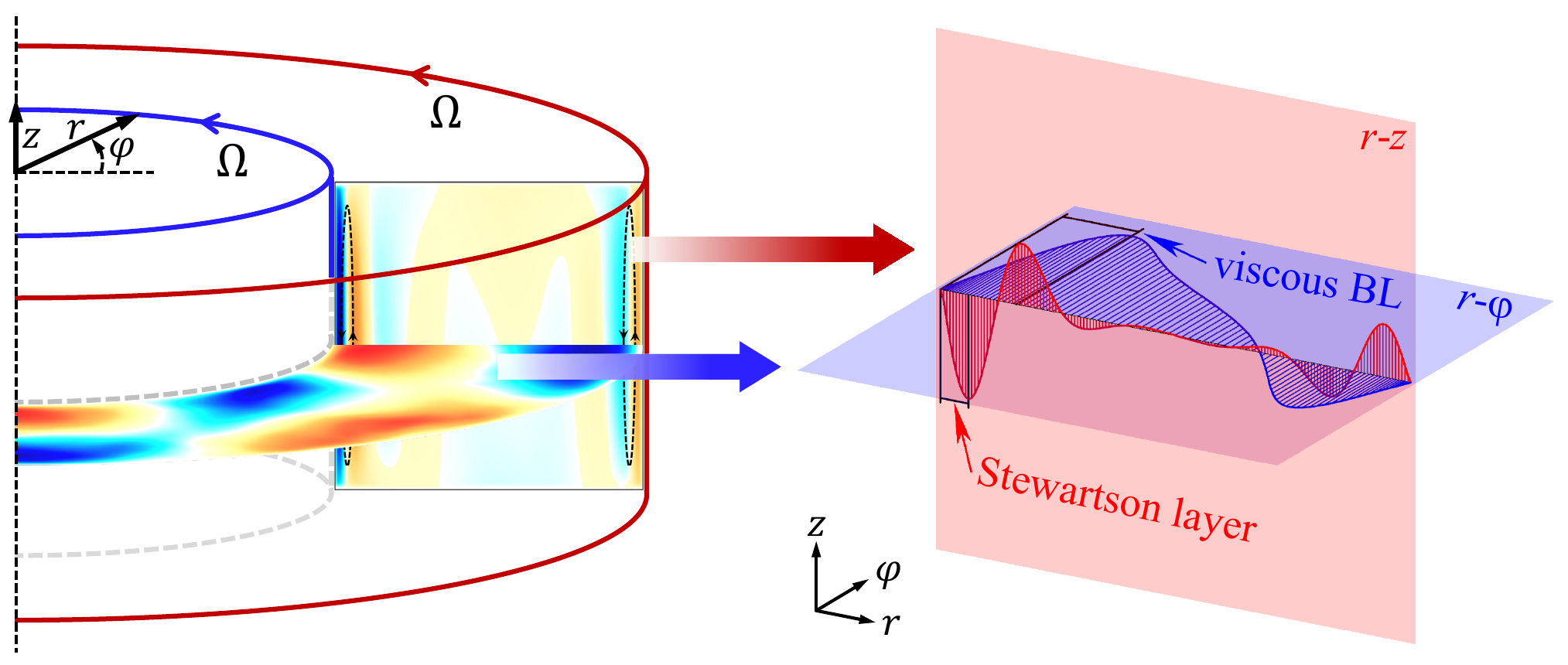}
\caption{\label{Fig2}Sketch of the formation of Stewartson layers in CC. Because of the strong rotation, viscous boundary layers develop near the inner (cooled) and outer (heated) cylinders in the $r-\varphi$ plane. The presence of Earth’s gravity causes two elongated vortex structures denoted by two circular dashed lines with arrows to form along the cylinders in the $r-z$ plane, leading to Stewartson layers near the cylinder walls. For the left diagram, the color contours represent the magnitude of the azimuthal velocity component in the $r-\varphi$ plane and axial velocity components in the $r-z$ plane, both obtained from numerical simulations. For the right diagram, the blue and red lines represent the azimuthal and axial velocity component profiles, respectively.}
\end{figure}

\textit{Theoretical explanation$-$}What is the unique aspect of CC compared to classical RBC? As shown in Fig.~\ref{Fig2}, the presence of Earth’s gravitational force breaks the top-bottom symmetric structure for pure centrifugal convection without gravity. Subsequently, two elongated vortex structures develop along the inner (cooled) and outer (heated) cylinders, leading to Stewartson layers that form near the cold and hot walls in the $r-z$ plane, in addition to the viscous boundary layer in the $r-\varphi$ plane. The thickness of the viscous boundary layer, $\delta_{\nu}$, is determined from the mainstream azimuthal velocity component, whereas the thickness of the Stewartson layer, $\delta_{st}$, is determined from the axial velocity component.

The recent work by \citet{Lai2025} showed that the dimensionless thickness of the Stewartson layer in CC, $\delta_{st}/L$, is governed by rotational effects, with $\delta_{st}/L \sim Ek^{1/3}$, consistent with results obtained in the RRBC system \cite{Stewartson1957, Stewartson1966, Kunnen2011, Kunnen2013}. Here, $Ek = \nu/(\Omega H^2)$ is the Ekman number, which characterizes the influence of the Coriolis force. Owing to the development of the viscous boundary layers and the Stewartson layers near the inner (cooled) and outer (heated) cylinders, their interaction may trigger the transition earlier than expected. In the following, we evaluate the properties of these two boundary layers.

First, the definition of $Ek$ and $Fr$ are
\begin{gather}
Ek = \frac{\nu}{\Omega H^2}, \quad Fr = \frac{\Omega^2 R_m}{g}.
\end{gather}
As the parameters $\nu, H, R_m, g$ are all constant in our experimental dataset of fixed $Pr$, both $Ek$ and $Fr$ only depend on $\Omega$; then we can obtain the practical relationship between $Ek$ and $Fr$ in the experiments:
\begin{gather}
Ek \sim Fr^{-1/2}.
\end{gather}
Hence, the thickness of the Stewartson layer $\delta_{st}$ follows
\begin{gather}
\delta_{st}/L \sim Ek^{1/3} \sim Fr^{-1/6}.\label{equ1}
\end{gather}

Next, in the RBC or CC system, according to $Re \sim Ra^{1/2}$, the thickness of the viscous boundary layer $\delta_{\nu}$ follows
\begin{gather}
\delta_{\nu}/L \sim Re^{-1/2} \sim Ra^{-1/4}.\label{equ2}
\end{gather}

Finally, assuming that the transition from the classical regime to the ultimate regime occurs when the thicknesses of the viscous boundary layer $\delta_{\nu}$ and the Stewartson layer $\delta_{st}$ are comparable in scale, we have:
\begin{equation}
\delta_{\nu}/L \sim \delta_{st}/L \Rightarrow Ra^{-1/4} \sim Fr^{-1/6} \Rightarrow Ra^* \sim Fr^{2/3}.
\end{equation}
which is in full agreement with the present experimental data as shown in Fig.~\ref{Fig1}(c). 

\begin{figure}[t]
\includegraphics[width=1.0\linewidth]{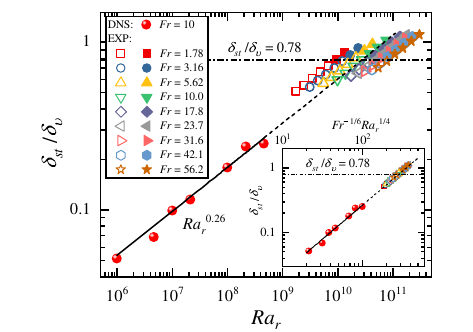}
\caption{\label{Fig3}Ratio $\delta_{st}/\delta_{\nu}$ of the thicknesses of the Stewartson layer to the viscous boundary layer, obtained from numerical simulations near the outer (heated) cylinder at a constant $Fr = 10$, as a function of $Ra_r$. The black solid line shows the fitted curves $\delta_{st}/\delta_{\nu} = 0.0021Fr^{-1/6}Ra_r^{0.26}$. The experimental points are obtained by substituting the $Fr$ and $Ra_r$ values from the experiment into this curve. The open and solid symbols represent the data points in the classical regime and the ultimate regime, respectively. Inset: the ratio $\delta_{st}/\delta_{\nu}$ as a function of $Fr^{-1/6}Ra_r^{1/4}$.}
\end{figure}

\textit{Physical mechanism$-$}We now propose a plausible explanation. When the thicknesses of the Stewartson layer and the viscous boundary layer coincide, their velocity components strongly couple near the inner and outer cylinders, forming a coupled flow that can distort the viscous boundary layer and enhance the anisotropy of local velocity variations. This interaction can trigger flow instability, eventually leading to the breakdown of the laminar boundary layer into turbulence. The turbulent boundary layer continuously disturbs the thermal boundary layer, characterized by a significant increase in heat transport. This mechanism is similar to that reported by \citet{King2009}, who found that the transition between rotationally dominated and nonrotating behaviour in RRBC is controlled by the relative thicknesses of the thermal (nonrotating) and Ekman (rotating) boundary layers.

To verify the above explanation, we perform three-dimensional direct numerical simulations that incorporate gravity and no-slip top and bottom boundaries, within a relatively low-$Ra_r$ range of $10^6 \sim 4.64\times10^8$. As an example, Fig.~\ref{Fig3} shows the ratio $\delta_{st}/\delta_{\nu}$, representing the thicknesses of the Stewartson layer $\delta_{st}$ determined from the axial velocity profile relative to that of the viscous boundary layer $\delta_{\nu}$ determined from the mainstream azimuthal velocity profile, obtained from numerical simulations near the outer (heated) cylinder at a fixed $Fr = 10$. It is observed that the ratio $\delta_{st}/\delta_{\nu}$ increases with $Ra_r$, following a power-law relation, i.e., $\delta_{st}/\delta_{\nu} = 0.0021Fr^{-1/6}Ra_r^{0.26}$. The scaling exponent of $0.26$ is close to the theoretical prediction $\delta_{st}/\delta_{\nu} \sim Ra^{1/4}$ (see Eq.~\ref{equ1} and Eq.~\ref{equ2}). Substituting the experimentally determined critical value $Ra_r^*$ at different $Fr$ values into the fitted relation yields critical ratios with a mean of 0.78, which is close to 1, as shown in Fig.~\ref{Fig3}. The inset of Fig.~\ref{Fig3} further presents $\delta_{st}/\delta_{\nu}$ as a function of $Fr^{-1/6}Ra_r^{1/4}$, where all data points collapse onto a single line, clearly demonstrating the universal character of this scaling relationship. This supports our theoretical explanation: a coupled flow structure, arising from the interaction of the Stewartson and viscous boundary layers, introduces distortions to the boundary layer flow, which in turn promotes flow instability and triggers the transition to turbulence.

\textit{Conclusion and outlook$-$}We present a combined experimental and numerical study of the transition from the classical regime to the ultimate regime in a supergravitational centrifugal convection system. Across a wide range of Froude numbers, the heat transport exponent increases from $\gamma = 0.27$ to $\gamma = 0.46$, marking the transition from the classical regime to the ultimate regime. The corresponding critical Rayleigh number for the onset of the ultimate regime is found to decrease with decreasing Froude number, highlighting the effect of Earth's gravity on the transition. The underlying mechanism can be understood as follows: In the classical regime, the Stewartson layer induced by Earth's gravity is thinner than the viscous boundary layer, so the latter controls boundary-layer stability. Once $Ra_r$ reaches the critical Rayleigh number, i.e., in the ultimate regime, the Stewartson layer becomes comparable to the viscous boundary layer ($\delta_{st}/\delta_{\nu} \approx 0.78$), and boundary-layer stability is then governed by their interaction. In this state, the velocity components are strongly coupled near the walls, potentially giving rise to a coupled flow that distorts the viscous boundary layer and enhances the anisotropy of local velocity variations, thereby triggering flow instability and causing the laminar layer to transition to turbulence. The turbulent boundary layer continuously disturbs the thermal boundary layer, characterized by a significant increase in heat transport. This interpretation is supported by extrapolated results from direct numerical simulations at relatively low $Ra_r$. Future work will need to directly measure and theoretically model the interaction of the boundary layers through experiments and simulations at the onset conditions, thereby providing direct evidence for this interpretation. Of course, the challenge for experiments is resolving the boundary layers at such high Rayleigh numbers, while for numerical simulations, it is reaching such high $Ra_r$, particularly in three-dimensional simulations to capture the additional effects of gravity. The current findings offer new insights into the transition from the classical to the ultimate regime in turbulent thermal convection.\\

\noindent
\textit{Acknowledgements$-$}We thank Detlef Lohse, Olga Shishkina and Jianjun Tao for their insightful discussions, comments, and suggestions and Yi-Bao Zhang, Feng Wang, and Zi-Hao Song for valuable discussions. This work is supported by the National Natural Science Foundation of China Excellence Research Group Program for Multiscale Problems in Nonlinear Mechanics (No.~12588201), the National Natural Science Foundation of China (No.~12502254), the New Cornerstone Science Foundation through the New Cornerstone Investigator Program and the XPLORER PRIZE, the China Postdoctoral Science Foundation (No.~2025M771852), the Postdoctoral Fellowship Program of CPSF (No.~GZC20251278), and the Shuimu Tsinghua Scholar Program (No.~2024SM299).\\

\noindent
\textit{Data availability$-$}The data that support the findings of this article are openly available \cite{data}.


\providecommand{\noopsort}[1]{}\providecommand{\singleletter}[1]{#1}%

\bigskip
\noindent
{\bf End Matter}\\

\noindent
\textit{Governing equations$-$}Dimensionless governing equations following the Oberbeck-Boussinesq assumption in a rotating reference frame can be expressed as
\begin{gather}
\nabla\cdot\bm{u}=0,\\
\frac{\partial T}{\partial t}+\nabla\cdot(\bm{u}T)=\frac{1}{\sqrt{Ra_r Pr}}\nabla^{2}T,\\
\nonumber
\frac{\partial\bm{u}}{\partial t}+\bm{u}\cdot\nabla\bm{u}=-\nabla p-Ro^{-1}\bm{\hat{e_z}}\times\bm{u} +\sqrt{\frac{Pr}{Ra_r}}\nabla^{2}\bm{u}\\
-\frac{2(1-\eta)}{(1+\eta)}rT\bm{\hat{e_r}}+\frac{Ra_g}{Ra_r}T\bm{\hat{e_z}},
\end{gather}
where $\bm{u}=(\bm{u_r},\bm{u_\varphi},\bm{u_z})$, $T$, $p$, $\bm{\hat{e_r}}$ and $\bm{\hat{e_z}}$ represent the velocity, temperature, pressure, and unit vectors in the radial and axial directions, respectively.

In these equations, there are four key control dimensionless parameters, namely the centrifugal Rayleigh number $Ra_r$, gravitational Rayleigh number $Ra_g$, inverse Rossby number $Ro^{-1}$, and Prandtl number $Pr$:
\begin{gather}
Ra_r = \frac{\alpha \Omega^2 R_m \Delta T L^3}{\nu \kappa}, \quad Ra_g = \frac{\alpha g \Delta T L^3}{\nu \kappa},\\
Ro^{-1} = 2\left[\frac{\alpha \Delta T R_m}{L}\right]^{-1/2}, \quad Pr = \frac{\nu}{\kappa},
\end{gather}
where $\alpha$, $\nu$, and $\kappa$ represent thermal expansion, kinematic viscosity, and thermal diffusivity, respectively. Here, $\Omega$ is the angular speed; $R_m=(R_o+R_i)/2$ is the distance between the center of the cell and the rotational axis; and $R_i$ and $R_o$ are the outer radius of the inner cylinder and the inner radius of the outer cylinder, respectively. $\Delta T$ is the temperature difference between the inner and outer cylinders. $L=R_o-R_i$ is the gap between the inner and outer cylinders. $\eta=R_i/R_o$ is the radius ratio.

One of the response dimensionless parameters is the Nusselt number $Nu$, which quantifies the ratio between the heat flux transported by the system $J_{\rm{total}}$ and that by conduction $J_{\rm{conduction}}$:
\begin{gather}
Nu = \frac{J_{\rm{total}}}{J_{\rm{conduction}}} = \frac{Q \mathrm{ln}(R_o/R_i)}{\lambda \Delta T 2\pi H},
\end{gather}
where $Q$ is the power input through the outer cylinder into the system, $\lambda$ is the thermal conductivity of the working fluid, and $H$ is the height of the outer cylinder.

The code utilized in this study is an extended development based on the open-source program AFiD \cite{verzicco1996, VanderPoel2015, Zhu2018}, which was originally designed for the classical RBC systems. In the CC system, we have incorporated the effects of the Coriolis force, gravity, and centrifugal buoyancy. No-slip and isothermal boundary conditions are imposed on the inner and outer cylindrical surfaces, while no-slip and adiabatic boundary conditions are applied to the top and bottom end covers. Grid generation employs a staggered grid scheme, with spatial derivatives solved via the second-order central difference method, and temporal derivatives processed by combining the third-order fractional-step Runge-Kutta method with Crank-Nicholson implicit terms. To ensure the temporal stability of the explicit scheme during integration, the Courant-Friedrichs-Lewy number is restricted to within 0.8 \cite{Courant1928}. Typically, the flow field reaches a statistically steady state after over 150 dimensionless time units of simulation, following which an additional 350 dimensionless time units of simulation are conducted to collect data for statistical analysis.

Previous studies have confirmed that the flow field in the CC system exhibits periodic characteristics in the circumferential direction \cite{Wang2022}. Thus, the circumferential simulation domain can be reduced from the full circumference to half or even a quarter of the circumference to reduce computational load. As long as the computational domain contains at least one pair of convective vortices, the heat transport and flow structures will be comparable to those from full-circumference simulations, without interfering with subsequent research \cite{Wang2022}. Thus, to save computing power, in most of the simulations presented in this paper, we adopt 1/4 of the circumference as the computational domain.\\

\noindent
\textit{Working fluid (Silicone oil-1cs)$-$}The advantage of using Silicone oil-1cs (denoted by open red squares, $Pr = 16.4$ at 25 $^\circ$C) is that, for the same range of temperature difference $\Delta T$, it can cover a wider range of $Ra_r$ around the critical $Ra_r^*$ from the classical to the ultimate regime than water (denoted by solid blue circles, $Pr=4.3$ at 40 $^\circ$C) or Novec-7200 (denoted by solid pink triangles, $Pr=10.3$ at 25 $^\circ$C) used in \citet{Jiang2020, Jiang2022}, as shown in Fig.~\ref{Fig4}. Table~\ref{tab1} lists the physical properties of Silicone oil-1cs at 25 $^\circ$C.\\

\noindent
\textit{Parameter space$-$}The parameter space explored in the $Fr$–$Ra_r$ and $Pr$–$Ra_r$ planes is shown in Figs.~\ref{Fig4}(a) and \ref{Fig4}(b), respectively. A total of 60 experiments were performed. The temperature difference $\Delta T$ ranged from 2.77 K to 21.35 K, and the Oberbeck-Boussinesq conditions are well satisfied \cite{Ahlers2006, Weiss2024}. The Rayleigh number $Ra_r$ spans $1.76\times10^{9}$ to $3.04\times10^{11}$, and the rotation rate ranges from 94 rpm to 528 rpm, corresponding to an effective gravity of 1.78 $g$ to 56.2 $g$.\\

\begin{table}[htbp]
\caption{\label{tab1}Physical properties of the Silicone oil-1cs at temperature 25 $^\circ$C.}
\begin{ruledtabular}
\begin{tabular}{cccccc}
\textrm{physical property}&\textrm{symbol}&\textrm{unit}&\textrm{Silicone oil-1cs}\\
\colrule
Thermal expansion    &$\alpha$     &$\rm{K}^{-1}$        & $1.29\times10^{-3}$\\
Kinematic viscosity  &$\nu$      &$\rm{m}^{2}\rm{s}^{-1}$& $1.00\times10^{-6}$\\
Thermal diffusivity  &$\kappa$   &$\rm{m}^{2}\rm{s}^{-1}$& $6.11\times10^{-8}$\\
Thermal conductivity &$\lambda$ &$\rm{W}(\rm{mK})^{-1}$  &        0.10        \\
Prandtl number       &$Pr$            & ... &                     16.4        \\
\end{tabular}
\end{ruledtabular}
\end{table}

\begin{figure}[htbp]
\includegraphics[width=1.0\linewidth]{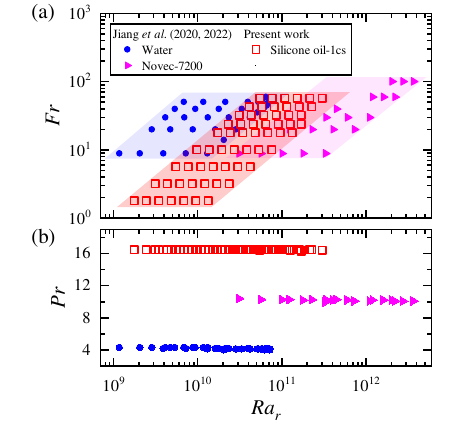}
\caption{\label{Fig4}(a) Explored parameter space in the $Fr$–$Ra_r$ plane. Open red squares denote the experimental data using Silicone oil-1cs as the working fluid in the present study. Experimental data using water (solid blue circles) and Novec-7200 (solid pink triangles) from \citet{Jiang2020, Jiang2022} are also shown. (b) Explored parameter space in the $Pr$–$Ra_r$ plane.}
\end{figure}

\end{document}